# Monopole polarization of $C_{60}$ fullerene shell

**M. Ya. Amusia[1,2] and A. S. Baltenkov[3]**

[1] *Racah Institute of Physics, the Hebrew University, Jerusalem, 91904 Israel*
[2] *Ioffe Physical-Technical Institute, St. Petersburg, 194021 Russia*
[3] *Arifov Institute of Ion-Plasma and Laser Technologies, Tashkent, 100125 Uzbekistan*

**Abstract**
We analyze using Poisson equation the spatial distributions of the positive charge of carbon atomic nuclei shell and negative charge of electron clouds forming the electrostatic potential of the $C_{60}$ fullerene shell as a whole. We consider also the case when an extra positive charge appears inside $C_{60}$ in course of e.g. photoionization of an endohedral A@C. We demonstrate that frequently used radial square-well potential $U(r)$ simulating the $C_{60}$ shell leads to nonphysical charge densities of the shell in both cases - without and with an extra positive charge inside. We conclude that the square well $U(r)$ modified by adding a Coulomb-potential-like term does not describe the interior polarization of the shell by the electric charge located in the center of the $C_{60}$ shell. We suggest another model potential, namely that of hyperbolic cosine shape with properly adjusted parameters that is able to describe the monopole polarization of $C_{60}$ shell. As a concrete illustration, we have calculated the photoionization cross-sections of H@$C_{60}$ taking into account the monopole polarization of the shell in the frame of suggested model. We demonstrate that proper account of this polarization does not change the photoionization cross-section.

## 1. Introduction

The idea that a phenomenological potential $U(r)$ formed by carbon atoms smeared inside a layer between two concentric spheres can describe electron interaction with fullerene $C_{60}$ is widely used despite the fact that this approach is an essential simplification of the real molecular field. Among the stepwise model potentials [1-3] perhaps the most popular is the spherical rectangular potential

$$U(r)=\begin{cases}-U_0, \text{if } r_0\le r\le r_0+\Delta;\\ 0, \quad \text{otherwise.}\end{cases} \tag{1}$$

Here $r_0$ denotes the inner radius, Δ - thickness, and $U_0$ - depth of $C_{60}$ potential well. A great number of theoretical investigations of electronic processes involving the fullerene $C_{60}$ itself, as well as endohedral atoms A@$C_{60}$ (see review [4] and references therein) employed this model potential.

Paper [5] analyzes the effect of interior static polarization of the $C_{60}$ shell in the process of endohedral A@$C_{60}$ photoionization. According to this paper "The quintessence of the effect is that the ion reminder $A^+$, ones the photoionization taken place and the photoelectron is produced, could polarize the $C_{60}$ cage…This causes that the fullerene shell potential $U(r)$ to be different versus the situation without consideration of static polarization". To take into account this effect the authors of [5] introduced a modified version of potential (1). Namely: they assumed that instead of (1) the potential acting upon electron that leaves the A@$C_{60}$ in course of its photoionization has the following form

$$U^*(r) = \begin{cases} \frac{\alpha}{r_0} - \frac{\alpha}{r_0+\Delta}, & \text{if } r \le r_0; \\ -U_0 + \frac{\alpha}{r} - \frac{\alpha}{r_0+\Delta}, & \text{if } r_0 \le r \le r_0+\Delta; \\ 0, & \text{otherwise.} \end{cases} \quad (2)$$

Here the parameter $\alpha$ is equal either to zero, $\alpha$=0 or to 1, if the static polarization is entirely ignored or complete included, respectively.

Investigated in [5] interior static polarization reflects, *per se*, spatial shifting of negative electron density of the $C_{60}$ shell relative to positive density of carbon ions. The positive electric charge of latter is smeared on the surface of the sphere with the radius *R*. Here *R* is the distance of the carbon atoms nuclei from the center of the $C_{60}$ shell. The shifting of electron density in each elementary volume of the $C_{60}$ shell under the action of positive atomic residue $A^+$ results in creating an induced electric dipole moment of this volume. The axes of all elementary dipole moments are directed to the center of the $C_{60}$ sphere and the electric component of the shell, as a whole, is shifted to the sphere center, which leads to the monopole polarization of the shell and in its turn changes a shape of the $C_{60}$ static potential.

In Section 2 we analyze with the help of Poisson equation the spatial distribution of the positive charge of the C-atomic nuclei and the negative charge of the electron clouds forming the electrostatic potential of the $C_{60}$ shell that is described by the formulas (1) and (2). In Section 3 we investigate the V-shaped (instead of stepwise-U-shaped) model potential $U(r)$ for the $C_{60}$ shell and charge distribution in it. In Section 4 we analyze the role of the $C_{60}$ monopole polarization in the photoionization of endohedral hydrogen atom H@$C_{60}$ within the framework of a new potential model described in Section 3. Section 5 presents our Conclusions.

## 2. Charge densities

The potential of electron interaction with $C_{60}$ Eq.(1) is connected with the potential of the electrostatic field $\varphi(r)$, in which the electron moves, by the relation $U(r) = -\varphi(r)$. Here the electron charge is equal to -1[*]. The Poisson equation defines the electrostatic field potential $\varphi(r)$

$$\Delta\varphi = -4\pi\rho, \quad (3)$$

where $\rho(r)$ is the density of the electric charges forming the spherically symmetric potential well (1). The following equation determines the radial dependence of this density

$$\frac{1}{r}\frac{d^2}{dr^2}[rU(r)] = 4\pi\rho(r). \quad (4)$$

[*] We employ the atomic units (at. un.) throughout the paper.

Using the Heaviside step function

$$\Theta(z) = [1 + \exp(z/\eta)]^{-1}\,, \tag{5}$$

let us rewrite the potential function (2) in the following form

$$U^*(r) = \left(\frac{\alpha}{r_0} - \frac{\alpha}{r+\Delta}\right)\Theta(r-r_0) - \left(U_0 - \frac{\alpha}{r} + \frac{\alpha}{r_0+\Delta}\right)\Theta(r_0-r)\Theta(r-r_0-\Delta)\,. \tag{6}$$

In Eq.(5) the diffuseness parameter $\eta$ is a fixed positive number that can be as small as we wish, and which can therefore ultimately be replaced by zero. The reasons why we replace the stepwise potentials (1) and (2) by diffuse potentials (6) will become clear later. Let us begin with the case α = 0 to understand what is the spatial distribution of the electric charge densities when the static polarization of $C_{60}$ shell is ignored. The numerical differentiation with the "Origin" packet is a suitable tool for dealing with Eq.(4).

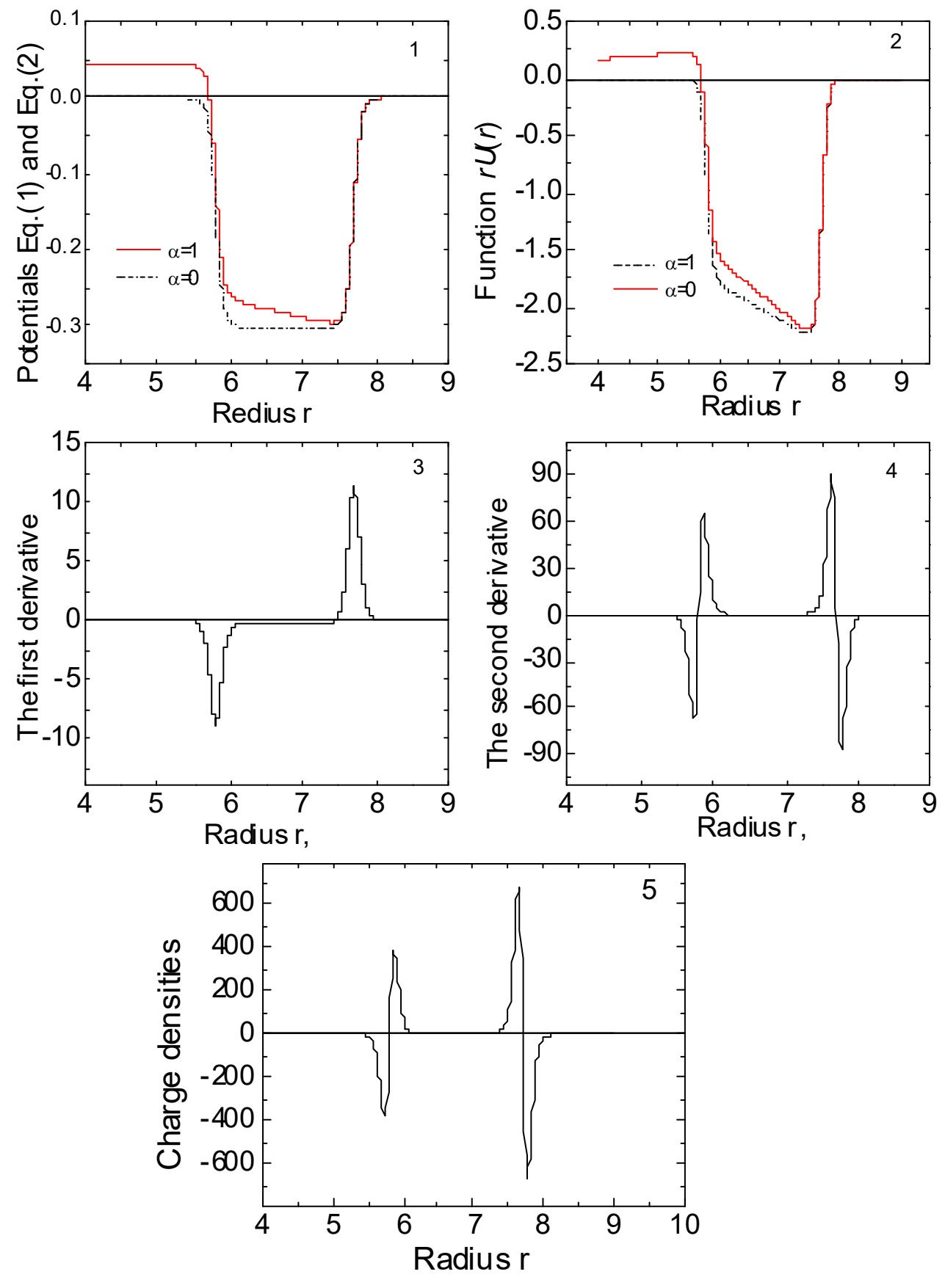

Fig. 1. The potential functions (1) and (2) (panel 1 and 2) and their first (panel 3) and second (panel 4) derivatives. Panel 5 is the charge distributions in the $C_{60}$ shell for α=0 and α=1. The parameters of potential wells are the same as in paper [2], namely $r_0$=5.8, Δ=1.9, $U_0$=0.301. The diffuseness parameter η=0.05. Both curves in panels 3-5 coincide with each other. All in *at. un.*

In figure 1 panels 1 and 2 present the potentials. One can see the first (panel 3) and the second (panel 4) derivatives of the function $rU(r)$. Panel 5 presents the charge distribution function $Q(r) = 4\pi r^2 \rho$. According to this figure, two concentric spheres with radiuses $r = r_0$ and $r_0+\Delta$, each with a double electric layers create the radial square-well potential (1). The thicknesses of these layers (for parameter η=0.05) are about 0.05. Both spheres are electrically neutral. On the surface of inner sphere about 36% of positive and negative charges are located. The rest charges of the $C_{60}$ shell are localized on the outer sphere surface. For potential well without diffuseness (η=0) the charge layers have zero thicknesses. The function $Q(r)$ is equal to zero everywhere except two points, $r = r_0$ and $r_0+\Delta$, in infinitesimal vicinities of which the charge densities are infinitely negative and positive.

Repeating the same procedure with potential function (2) (α=1) we come to the same charge distribution as in case α=0 (see panel 5 of figure 1). The reason for this, at the first glance unexpected, result is as follows. Let us apply the Laplacian Δ from the Poisson equation (3) to additional terms in Eq.(2).

For the first line we have

$$\Delta\left(\frac{\alpha}{r_0}-\frac{\alpha}{r_0+\Delta}\right)\equiv 0. \tag{7}$$

For the second line one has

$$\Delta\left(\frac{\alpha}{r}-\frac{\alpha}{r_0+\Delta}\right)=-4\pi\alpha\delta(\mathbf{r}), \tag{8}$$

since, as is well-known, the Coulomb potential $\alpha/r$ is the Green function for the Poisson equation [6].

Again, we have zero in the right side of Eq.(8) because $\mathbf{r}\neq 0$ in this line. Thus, the additional terms in potential function (2) do not describe changes in the mutual disposition of electric charges in the $C_{60}$ shell, as well as static monopole polarization of the fullerene shell by the additional electric charge located in the center of the shell. Thus, the transition from (1) to (2) does not include at all the charge redistribution under the action of photoionization of atom A in A@$C_{60}$, contrary to the assumption made in [5].

In the next section we will present an example of a model potential, the parameters variation of which will describe the monopole polarization of $C_{60}$ shell.

**2. Model potential with hyperbolic cosine**

The following requirements guide us in the selection of the model potential well $U(r)$ that properly describes $C_{60}$ shell. The potential $U(r)$ has to be attractive and an $s$-level should exist in it with the binding energy equal to $E_s^- \cong -2.65eV$ that is the experimental value of the electron affinity energy of $C_{60}^-$. The $p$-like bound state can be considered as a ground state provided that the extra electron interaction with the field of electromagnetic radiation is neglected. The function $U(r)$ should be localized in a rather thin spherical layer with the thickness $\Delta$ of about few atomic units in the vicinity of the fullerene radius $R$. As shown in [7], in order to avoid the unphysical splitting of positive charge of the $C_{60}$ shell into the two concentric spheres (see figure 1) we have to find among the different potential functions $U(r)$ a potential well with non-flat bottom. In addition, the function $U(r)$ should exponentially decrease with radius $r$ as a potential of any neutral atomic-like system.

It is evident that the number of such potentials is unlimited. Let us consider one of them, namely the cosh -bubble potential family [8]

$$U(r)=-\frac{U_{\max}}{\cosh^n[\beta(r-R)]}, \tag{9}$$

that was called so in analogy with the Dirac-bubble potential [9]

$$U(r)=-U_0\delta(r-R). \tag{10}$$

The function (9) exponentially decreasing with radius $r$ and obeys all above-mentioned requirements. In further consideration we choose for simplicity in Eq.(9) the parameter $n$=1. In the middle of the maximal depth of the well (9), the thickness of potential well $\Delta$ is connected with the parameter $\beta$ by the following relation

$$\Delta = \frac{2}{\beta}\ln(2+\sqrt{3}) = 2.633916/\beta . \quad (11)$$

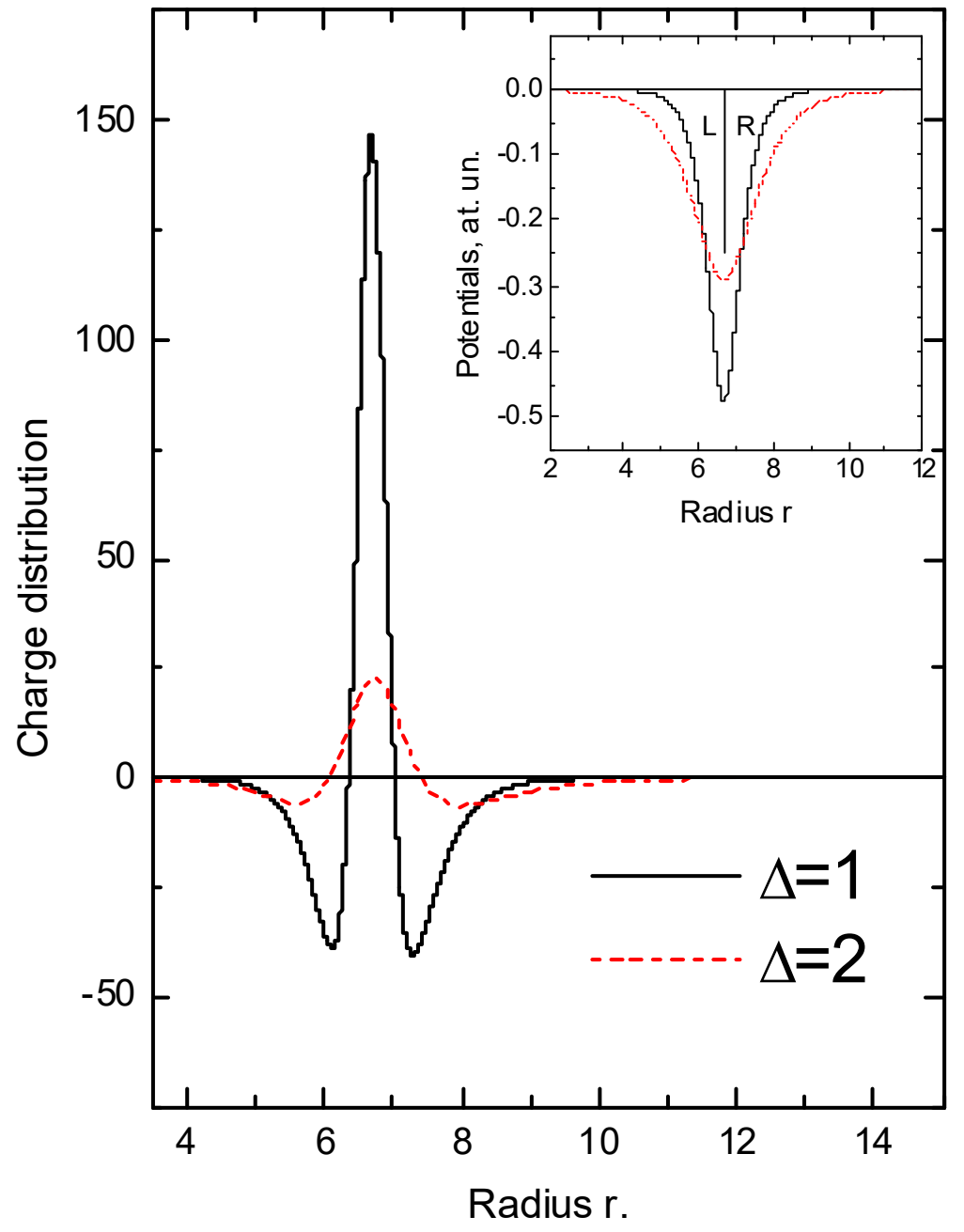

Fig. 2. Charge distribution for potential functions Eq.(9); insert is the Cosh-bubble potential wells; letters $L$ and $R$ are the left and right "cheek-bones" of potential Eq.(9). The parameters of the wells are: for thickness $\Delta$=1, depth $U_{\max}$=0.4762; for thickness $\Delta$=2, depth $U_{\max}$=0.2898; radius of potential wells is $R$=6.665; all in *at. un.*

The two parameters of this potential $U_{\max}$ and $\Delta$ are connected in such a way that in the potential well (9) there exists an $s$ state with specified above energy $E_s$=-2.65 eV (for details see [8]).

Let us apply the Poisson equation (4) to potential function (9). Repeating our actions done before, we obtain the spatial electric charge distribution that produces the potential well (9). Figure 2 presents this charge distributions for potentials with thicknesses $\Delta$=1 and 2. The charge distributions in this figure fundamentally differ from that in figure 1. The charge density in figure 2 is a three-layer sandwich, where the middle layer represents positively charged $C^{4+}$ ions. The inner and outer layers represent negatively charged clouds of collectivized 240 electrons of $C_{60}$. The total charge of the shell (9) is equal to zero because the cosh-bubble potential (9) is a short-range potential. In the case $\Delta$=1 about 45% of the negative charge is located in the inner electronic cloud. The rest negative charge of $C_{60}$ shell is localized in the outer electronic cloud. In the case $\Delta$=2 about 40% of negative charge are in the inner cloud.

Let us show that changing the left "cheek-bone" of potential well (9) relative to the right one corresponds to transition of the part of collectivized electrons from the outer electron cloud to the inner one and *vice versa*, i.e. this changing describes the monopole polarization of the $C_{60}$ shell. Using the step function (5), we replace the constant thickness $\Delta$ in Eq.(9) by the following expression

$$\Delta(r) = \Delta_L + (\Delta_R - \Delta_L)\Theta(R - r) . \quad (12)$$

Here $\Delta_L$ and $\Delta_R$ are the thicknesses of the left and right "cheek-bones" of potential well (9), respectively. Applying again Eq.(4) to the potential function (9) and taking into account Eq.(12), we obtain the charge distributions in the $C_{60}$ shell for the set of $\Delta_L$ and $\Delta_R$ parameters.

The evolution of the negative charge distribution in the left and right sides of electron clouds we see in figure 3. Solid black lines in this figure are the electronic spatial distribution with no monopole polarization of the $C_{60}$ shell ($\Delta_L$=$\Delta_R$=$\Delta$). Dash-dot magenta lines are electronic charges for the maximally considered differences of the potential well thicknesses: $\Delta_L$=0.7 and $\Delta_R$=1.3 in the upper panel and $\Delta_L$=1.7 and $\Delta_R$=2.3 in the lower one. These sets of the left ($\Delta_L$) and right ($\Delta_L$) thicknesses correspond to the maximal considered shifts of electron clouds relative the positive charges of the shell. All other curves between black and magenta lines correspond to the partial monopole polarization of the $C_{60}$ shell. Comparison of the areas under curves for inner and outer clouds shows that the charge of the inner cloud smoothly increases from 45% up to 60% in the upper panel and from 40% to 50% in the lower one when shell polarization become stronger.

In the next Section we will apply the potential (9) to calculate the photoionization of hydrogen atom located at the center of the $C_{60}$ shell.

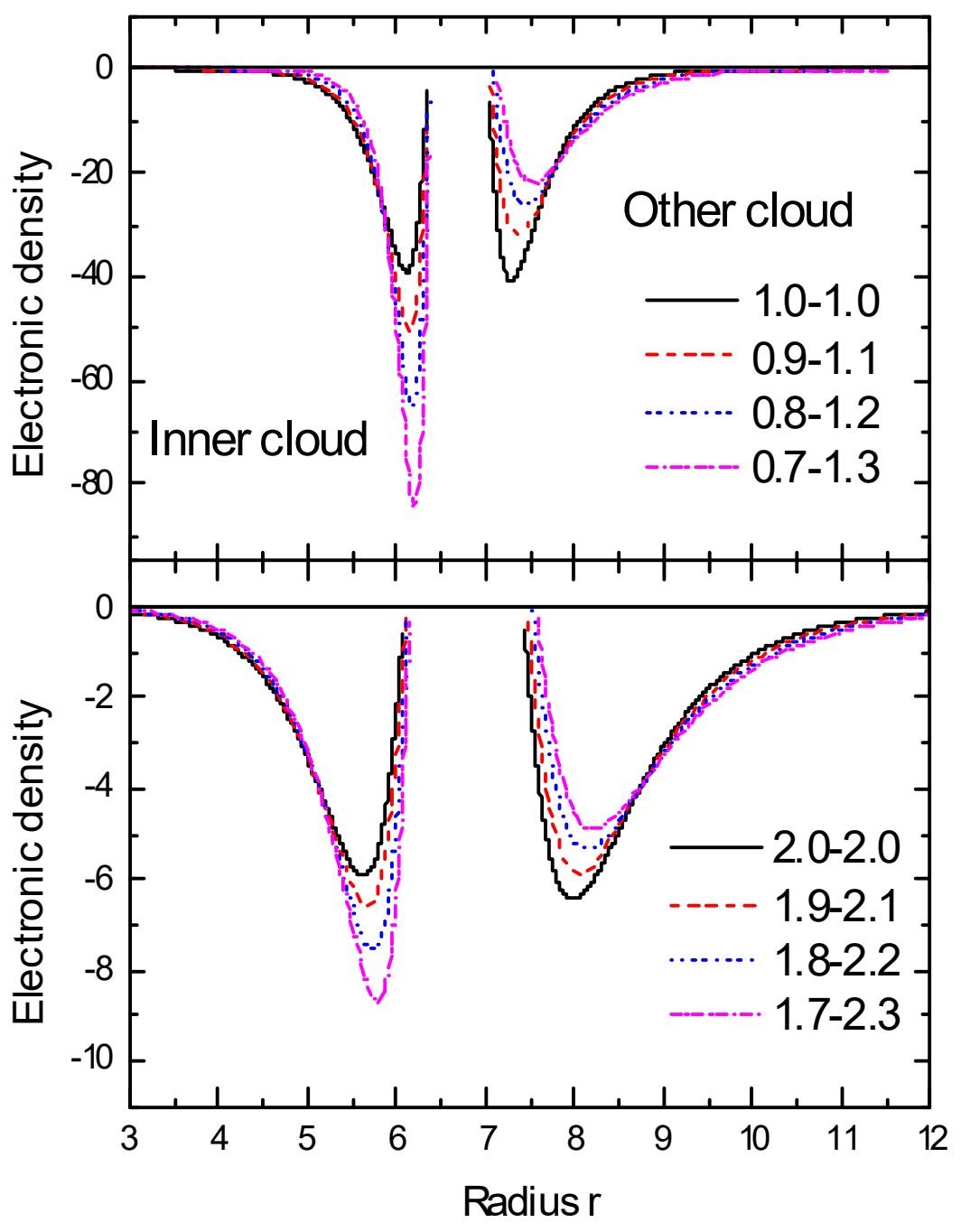

Fig. 3. Spatial distributions of negative charges in the inner and outer electronic clouds for a number of combinations of the cosh -bubble potential thicknesses. In the upper panel: the line 1.0-1.0 corresponds to the combination $\Delta_L$=1.0 and $\Delta_R$=1.0; the line 0.9-1.1 to combination $\Delta_L$=0.9 and $\Delta_R$=1.1; *etc*. The same is for the lower panel. The middle part of curves (the positive charge distribution) is almost the same for all combinations of thicknesses. All in *at. un.*

### 3. Photoionization of an endohedral hydrogen atom

The specific feature of an endohedral atom photoionization is presence of oscillations in the photoionization cross-sections, commonly known as confinement resonances. They were observed in the photoionization of endohedral Xe atom in Xe@$C_{60}$ molecule [10]. Figure 4 present the photoionization cross-sections of an endohedral hydrogen atom H@$C_{60}$ calculated with formulas (9) and (10) for potentials of the $C_{60}$ shell. As we can see, the cosh -bubble potential exhibits confinement resonances. The amplitudes of the confinement oscillations resulting from the cosh -bubble potential are somewhat smaller than those of the Dirac-bubble potential (10). Quite naturally, due to diffuseness the potential (9) somewhat smear the confinement resonances in the photoionization cross-sections (especially for $\Delta$=2), but do not eliminate them. Coincidence of curves for $\Delta$=1 and combination 0.7-1.3 of thicknesses (corresponding to maximal considered shift of the electronic charge in the $C_{60}$ shell) demonstrates the absence of the monopole polarization effect on the photoionization of endohedral hydrogen atom. Thus, it is emphatically incorrect that interior static polarization of the $C_{60}$ shell "may not be ignored in the photoionization of endohedral atoms near threshold" [5].

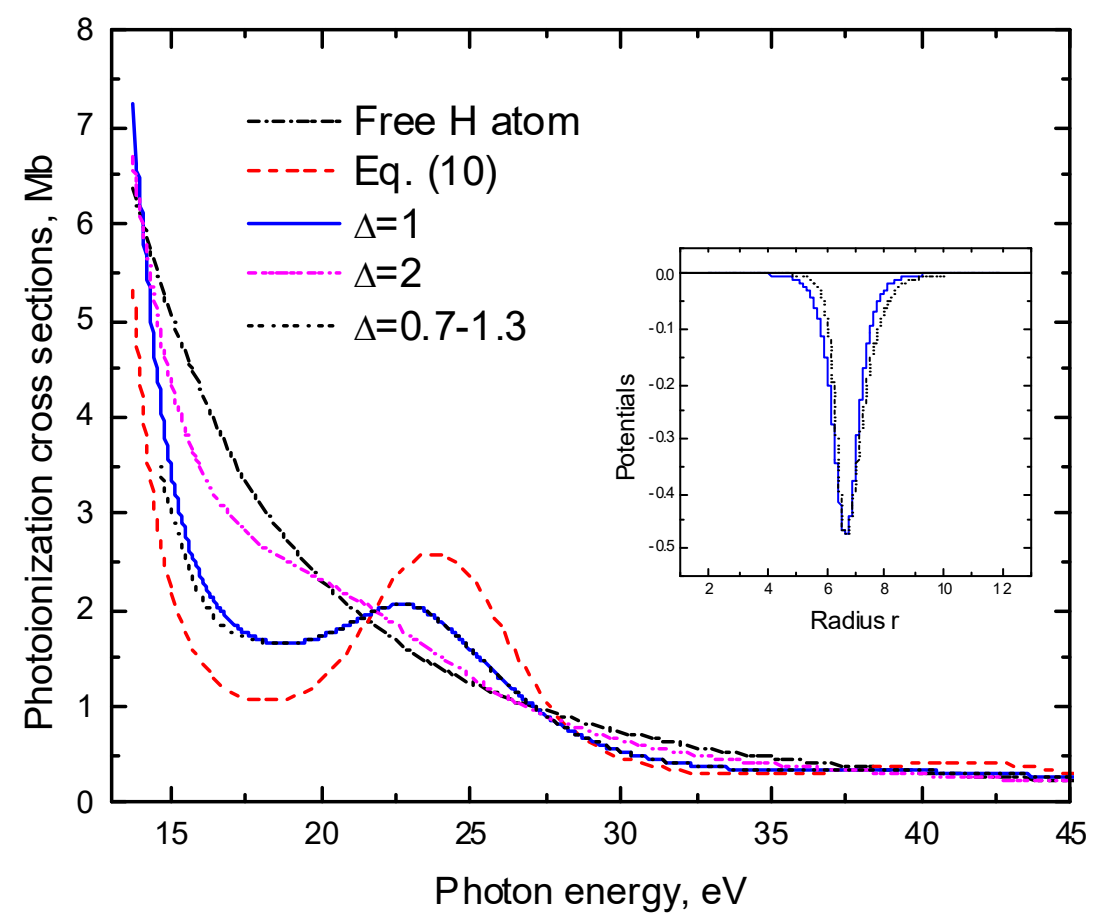

Fig. 4. Photoionization cross-sections of the ground state of the H atom in H@$C_{60}$ for parameters $\Delta$=1 and 2 of cosh -bubble potential (9) (solid blue and dash-dot-dot magenta lines); black dot line corresponds to $\Delta_L$=0.7 and $\Delta_R$=1.3 combination of the cosh -potential thicknesses. The cross-section resulting from a Dirac-bubble potential Eq.(10) is dashed red curve. The cross-section for the free H atom is dash-dot-dot black curve.
Insert shows the shapes of potential well (9) with $\Delta_L$=0.7 and $\Delta_R$=1.3 parameters, all in *at. un.*

Described in paper [5], changes in the cross sections are due to the introduction into the square-well potential model Eq.(1) additional arbitrary parameters that have nothing to do with changes in the mutual disposition of electric charges in the shell under the action of inner atomic core electric charge.

## 4. Conclusions

We have analyzed the spatial distribution of the positive charges of atomic nuclei and the negative charges of the electron clouds forming the electrostatic potential of $C_{60}$, as a whole. We demonstrated that the square-well potential (1) corresponds to onion-type structure for $C_{60}$ fullerene shell, with two concentric spheres and a gap $\Delta$ between them, see figure 1. Such charge distribution in the $C_{60}$ shell is in contradiction with the real structure of the $C_{60}$ molecule. The modification of formulas for the square-well potential (1) by means of addition to it the Coulomb-potential-like terms Eq.(2) does not describe the interior polarization of the shell by the electric charges located in the center of the $C_{60}$ shell.

The phenomenological potentials simulating the $C_{60}$ shell potential, if they are generated by a physically reasonable three-layer charge density (see figure 2), should belong to a family of potentials with a non-flat bottom. We propose and discuss potential (9) with hyperbolic cosine as a model of the $C_{60}$ shell. We demonstrated that the monopole polarization of $C_{60}$ shell by an extra inner electric charge is described by the parameters variation of this model potential. We calculated the photoionization cross-sections of an endohedral hydrogen atom H@$C_{60}$ with this potential and demonstrated that cosh -bubble potential exhibit confinement resonances as well. Furthermore, our analysis demonstrates that monopole polarization of the $C_{60}$ shell by atomic residue A+ in the center of shell has no effect on the amplitudes of confinement oscillations, contrary to conclusions made in [5].

The problem with model description of the shape and parameters of the fullerene shell potential is similar, to some extent, to that in nuclear physics where, in addition, the potential for nucleon-nucleon interaction is unknown. To describe the magic nuclei, the researchers selected complicated shapes of the potentials, such as, for example, the “Elsasser wine bottle” or “Mexican hat” potentials, that depend on a great number of parameters. In such a way one could model all the magic numbers of nuclei [12, 13]. We have to hope that more detailed experimental investigations of $C_{60}$ itself and endohedral systems A@$C_{60}$ will discover a new avenue to modification of the $C_{60}$ shell models.

**Acknowledgments**
ASB is grateful for the support to the Uzbek Foundation Award OT-Ф2-46**.**

**References**

1. M. J. Puska and R. M. Nieminen, Phys. Rev. A **47** 1181 (1993).
2. Y. B. Xu, M. Q. Tan and U. Becker, Phys. Rev. Lett. **76** 3538 (1996).
3. W. Jaskólski, Phys. Rep. **271** 1 (1996).
4. V. K. Dolmatov, in *Theory of Confined Quantum Systems: Part Two*, edited by J. R. Sabin and E. Brändas, Advances in Quantum Chemistry (Academic Press, New York, 2009), vol. 58, pp. 13-68.
5. V. K. Dolmatov and S. T. Manson, Phys. Rev. A **82**, 023422 (2010).
6. L. D. Landau & E. M. Lifshitz, *The Classical Theory of Fields* (Pergamon Press 1971).
7. A. S. Baltenkov, S. T. Manson and A. Z. Msezane, J. Phys. B **48** 185103 (2015).
8. A. S. Baltenkov, e-print arXive: 1406.2165
9. L. L. Lohr and S. M. Blinder, Chem. Phys. Lett. **198** 100 (1992).
10. R. A. Phaneuf, *et al*, Phys. Rev. A **88** 053402 (2013).
11. A. S. Baltenkov, Phys. Lett. A **254** 203 (1999).
12. M. G. Mayer and H. D. Jensen, *Elementary Theory of Nuclear Shell Structure* (New York: Wiley, 1955).
13. E. Feenberg, *Shell Theory of the Nucleus* (Princeton, NJ: Princeton University Press, 1955).

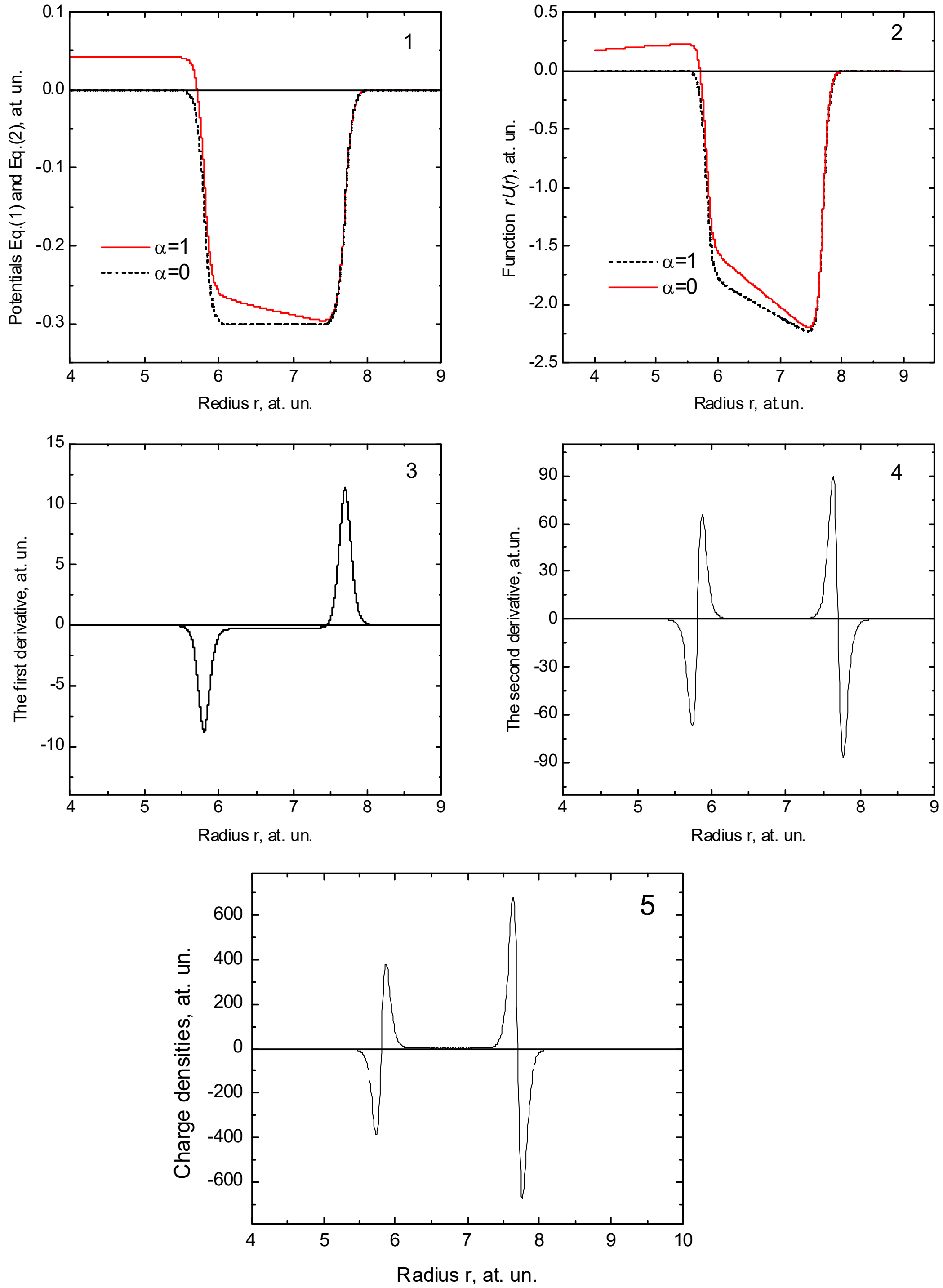

Fig. 1. The potential functions (1) and (2) (panel 1 and 2) and their first (panel 3) and second (panel 4) derivatives. Panel 5 is the charge distributions in the $C_{60}$ shell for α=0 and α=1. The parameters of potential wells are the same as in paper [2], namely $r_0$=5.8, Δ=1.9 at. un., $U_0$=0.301 at. un. The diffuseness parameter η=0.05. Both curves in panels 3-5 coincide with each other. All in *at. un*.

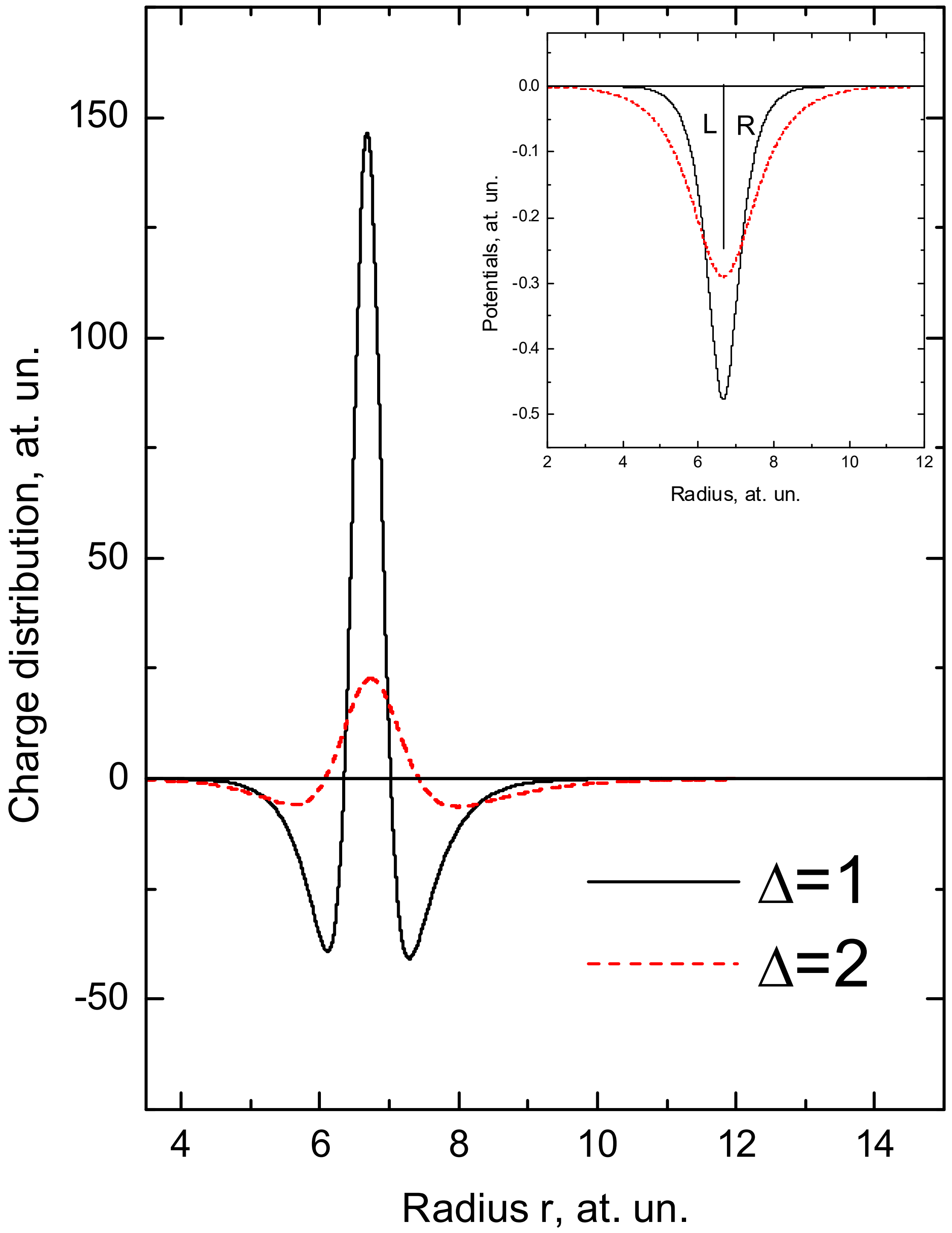

Fig. 2. Charge distribution for potential functions Eq.(9); insert is the Cosh-bubble potential wells; letters *L* and *R* are the left and right "cheek-bones" of potential wells Eq.(9). The parameters of wells: for thickness $\Delta$=1, depth $U_{max}$=0.4762; for thickness $\Delta$=2, depth $U_{max}$=0.2898; radius of potential wells $R$=6.665; all in at. un.

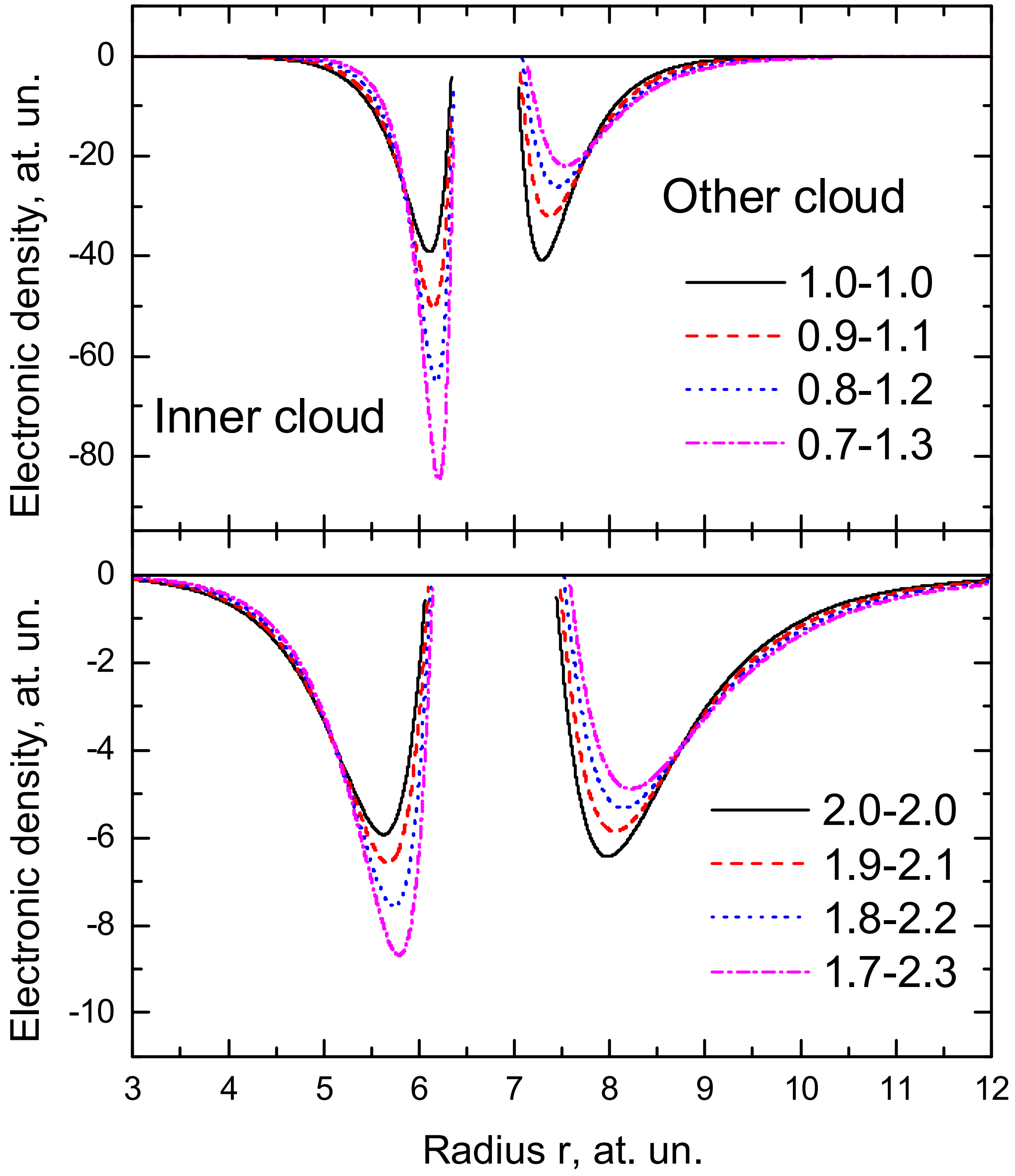

Fig. 3. Spatial distributions of negative charges in the inner and outer electronic clouds for the following combinations of the cosh-bubble potential thicknesses. In the upper panel: the line 1.0-1.0 corresponds to the combination $\Delta_L$=1.0 and $\Delta_R$=1.0; the line 0.9-1.1 to combination $\Delta_L$=0.9 and $\Delta_R$=1.1; *etc*. The same is for the lower panel. The middle part of curves (the positive charge distribution) is almost the same for all combinations of thicknesses. All in *at. un*.

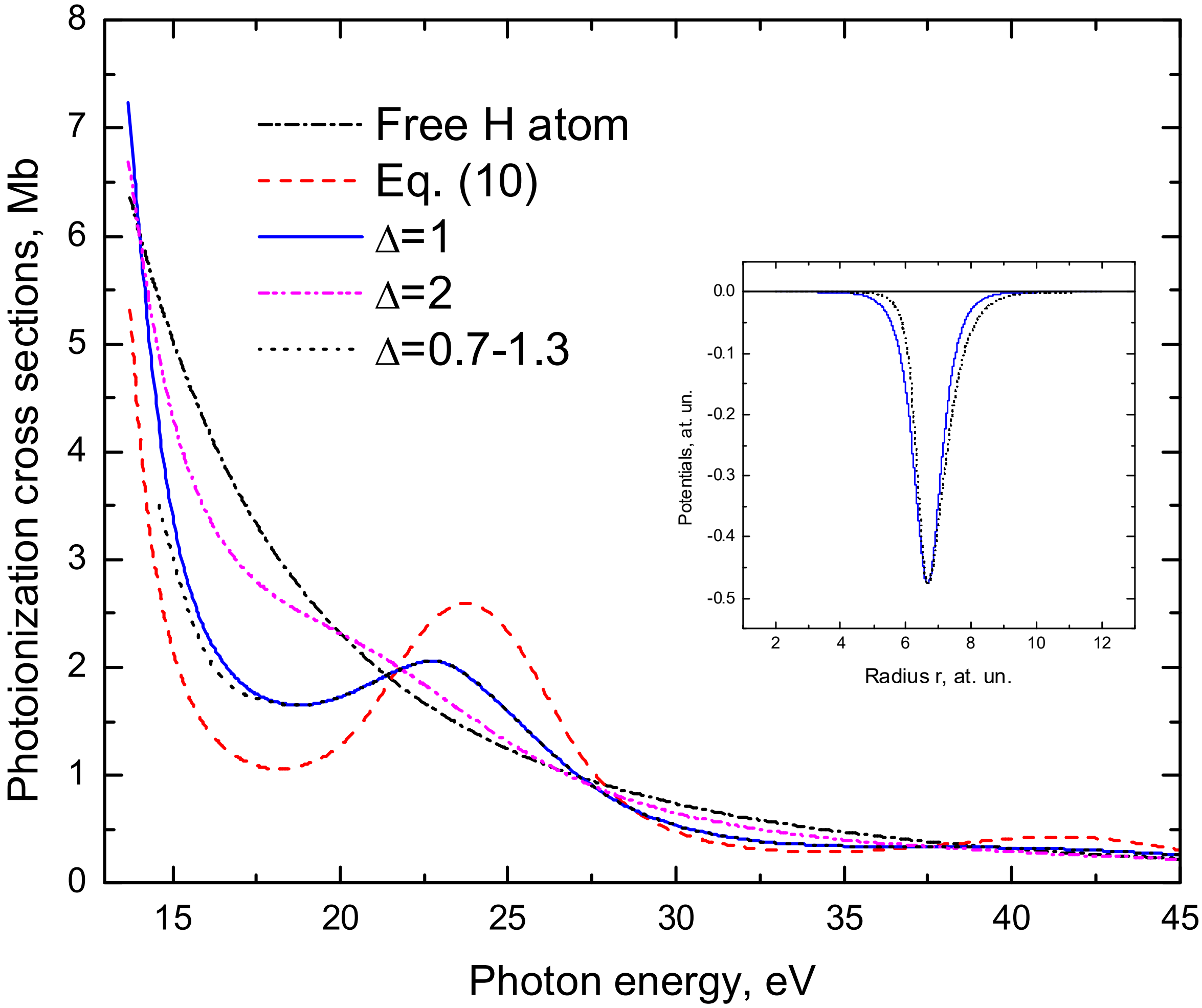

Fig. 4. Photoionization cross-sections of the ground state of the H atom in H@$C_{60}$ for parameters Δ=1 and 2 of cosh-bubble potential (9) (solid blue and dash-dot-dot magenta lines); black dot line corresponds to $\Delta_L$=0.7 and $\Delta_R$=1.3 combination of the Cosh-potential thicknesses. The cross-section resulting from a Dirac-bubble potential Eq.(10) is dash red curve. The cross-section for the free H atom is dash-dot-dot black curve.
Insert shown the shapes of potential well (9) with $\Delta_L$=0.7 and $\Delta_R$=1.3 parameters, all in *at. un*.